\documentclass[11pt,letterpaper]{article} 
\usepackage{amsmath,amssymb}
\usepackage{graphicx}
\usepackage[pdftex]{hyperref}

\usepackage[left=2.0cm, right=2.0cm, top=3.0cm, bottom=3.0cm]{geometry}

\title{\textbf{Inflation driven by a holographic energy density}}

\author{A.~Oliveros, Mario A.~Acero \\
{\small Programa de F\'isica, Universidad del Atl\'antico Carrera 30 N\'umero 8-49 Puerto Colombia-Atl\'antico, Colombia}
}

\begin{document}

\maketitle

\abstract{In this letter we study  a model of inflation in which the inflationary regimen comes from a 
type of holographic energy density. In particular, we consider the Granda-Oliveros proposal for the holographic energy density,
which contains two free dimensionless parameters, $\alpha$ and $\beta$. This holographic energy density is associated to the so-called Granda-Oliveros infrared cutoff (G-O cutoff). Additionally, since in the inflationary regimen the energy
scales are very high, it is necessary to modify the G-O cutoff taking into account a correction due to the ultraviolet cutoff.
In this way, we obtain an algebraic equation which implicitly includes the Hubble parameter (as a function of e-folding number, $N$) and
from this, we calculate the Hubble slow-roll parameters and the values of the inflationary observables: the scalar spectral index of the curvature perturbations and its running, the tensor spectral index and the tensor-to-scalar ratio. Finally, since the values for these inflationary observables are known (Planck 2018 observations), we present constraints on the parameters $\alpha$ and $\beta$ to make this a viable model.}

\section{Introduction}
The holographic principle is considered as a new paradigm in quantum gravity and was first put forward by
't Hooft \cite{thooft} in the context of black hole physics and later extended by Susskind \cite{susskind} to string theory; however this
idea has its roots in the investigation on black hole thermodynamics performed by Bekenstein \cite{bekenstein} and Hawking \cite{hawking} in the 70's.
According to the holographic principle, the entropy of a system scales not with its volume, but with its surface area
\cite{bousso}. In other words, the degrees of freedom of a spatial region reside not in the bulk but only at the boundary
of the region, and the number of degrees of freedom per Planck area is no greater than unity. In reference \cite{cohen}, it was suggested that, in quantum field theory, a short distance cutoff is related to a long distance cutoff due to the limit set by the formation
of a black hole, namely, if $\rho$ is the quantum zero-point energy density caused by a short distance cutoff,
the total energy in a region of size $L$ should not exceed the mass of a black hole of the same size, thus
$L^3\rho\leq LM_{pl}^2$. The largest (infrared cutoff) $L_{IR}$ allowed is the one saturating this inequality, thus
\begin{equation}\label{eq1}
\rho=3c^2M_{pl}^2L_{IR}^{-2},
\end{equation}
where $c$ is an arbitrary parameter, and $M_{pl}$ is the reduced Planck mass.\\

In the last years, the holographic principle has been widely studied in the cosmological context, mainly to solve the dark energy 
problem. In this case the infrared cutoff has a cosmological origin and the energy density given by Eq.~(\ref{eq1}) is called holographic dark energy (see \cite{li} for an extensive review).
In the literature some infrared cutoffs for the holographic dark energy density have been proposed. For example, the Hubble length \cite{hsu},
the particle horizon or the future event horizon \cite{li2}, the generalized holographic dark energy where the infrared cutoff is identified with some combination of the FRW parameters (Hubble constant, particle and future horizons, cosmological
constant and universe life-time -if finite-) \cite{odintsov,odintsov1},
the age of the universe or the conformal time \cite{cai, wei}, 
the inverse square root of the Ricci scalar curvature \cite{gao}, the G-O cutoff \cite{oliveros1, oliveros2}, among others. In this work we will use the G-O cutoff in an inflationary context. This cutoff, in addition to the square of the Hubble parameter, also contains the time derivative of the Hubble parameter, namely
\begin{equation}\label{eq2}
L_{IR}=\frac{1}{\sqrt{\alpha H^2+\beta\dot{H}}},
\end{equation}
where $\alpha$ and $\beta$ are arbitrary dimensionless parameters. This cutoff was proposed by the authors of Refs.~\cite{oliveros1, oliveros2}, considering  purely dimensional reasons. In addition to the fact that the underlying origin of the holographic dark energy (or holographic inflation) is still unknown, the inclusion of the time derivative of the Hubble parameter may be expected as this
term appears in the curvature scalar \cite{gao}. On the other hand, contrary to the infrared cutoff given by the future event horizon \cite{li2}, this model avoids the causality problem. 
Recently, in \cite{odintsov2}  the holographic principle was used for first time in the inflationary context taking the
particle or future event horizons as infrared cutoff and adding a simple correction due to the ultraviolet cutoff, whose role is non-negligible at the high energy scales of inflation.
Additionally, in that scenario, the authors have calculated the Hubble slow-roll parameters and the inflation-related observables, such as the scalar spectral index and its running, the tensor-to-scalar ratio, and the tensor spectral index, which are in agreement with the Planck 2018 observations \cite{planck}. Also, the authors suggest that their formalism can be used with other generalized infrared cutoffs and
they show that, in principle, is possible to obtain a correspondence between the generalized holographic energy densities and some geometrical inflationary models, such as $F(R)$ gravity \cite{odintsov3}, Gauss-Bonnet and $f(G)$ inflation \cite{barrow, kanti}, $f(T)$ inflation, etc.\\

Following the above ideas, in this work we use the G-O proposal for the holographic energy density as the source of inflation. We will also take into account a correction due to the ultraviolet cutoff which modify our infrared cutoff. In this way, we proceed to obtain  the Hubble parameter, the Hubble slow-roll parameters and the values of the inflationary observables: the scalar spectral index of the curvature perturbations and its running, the tensor spectral index and the tensor-to-scalar ratio. Finally,  since the values for these inflationary observables are known \cite{planck}, then we obtain  the constraints for the values of the parameters $\alpha$  and $\beta$ that make it a viable model.\\

\section{Inflation with G-O infrared cutoff}
In this section we consider the G-O holographic energy density as source of inflation.  To this aim, the first step is to consider a flat, homogeneous and isotropic universe whose metric  is given by the Friedmann-Robertson-Walker (FRW) metric:
\begin{equation}\label{eq3}
ds^2=-dt^2+a(t)^2 \delta_{ij}dx^idx^j,
\end{equation}
where $a(t)$ is the scale factor.\\

Introducing the metric Eq.~(\ref{eq3}) in the usual Einstein equations and considering the (00)-component, we obtain
the first Friedmann equation:
\begin{equation}\label{eq4}
3M_{pl}^2H^2=\rho;
\end{equation}
in this case $\rho$ represents the holographic energy density given by Eq.~(\ref{eq1}) which drives the inflation, and it is assumed that in the early universe there are not additional energy densities contributing in the right side of Eq.~(\ref{eq4}), such as matter and radiation.\\

In addition to the G-O proposal for the holographic energy density as source of the inflation, we also take into account a correction due to the ultraviolet cutoff $\Lambda_{UV}$ whose role is non-negligible at the high energy scales of inflation \cite{odintsov2, odintsov4}. This correction modifies the infrared cutoff in Eq.~(\ref{eq2}) as follows:
\begin{equation}\label{eq5}
\tilde{L}_{IR}=\sqrt{L_{IR}^2+\frac{1}{\Lambda_{UV}^2}}=\sqrt{\frac{1}{\alpha H^2+\beta\dot{H}}+\frac{1}{\Lambda_{UV}^2}}.
\end{equation}
Now, replacing the new infrared cutoff given by Eq.~(\ref{eq5}) in Eq.~(\ref{eq1}), the Friedmann equation (\ref{eq4})
takes the form
\begin{equation}\label{eq6}
H^2=\frac{\Lambda_{UV}^2(\alpha H^2+\beta\dot{H})}{\alpha H^2+\beta\dot{H}+\Lambda_{UV}^2},
\end{equation}
where the parameters $\alpha$ and $\beta$ are playing role of the parameter $c$. In order to simplify the forthcoming calculations, we make the change of variable $N=\ln{(a/a_{i})}$, from which $dN=Hdt$, where $N$ is the
e-folding number and $a_i$ is the scale factor at the beginning of inflation. Furthermore, this change of variable facilitates
the computation of the inflationary observables that we will consider later in this work. Thereby, taking into account that
$\dot{H}=\frac{1}{2}\frac{d}{dN}H^2(N)$, Eq.~(\ref{eq6}) is reduced to a more suggestive form, namely
\begin{equation}\label{eq7}
\frac{d}{dN}H^2(N)=\frac{2H^2(N)\left[\Lambda_{UV}^2(1-\alpha)+\alpha H^2(N)\right]}{\beta\left[\Lambda_{UV}^2-H^2(N)\right]},
\end{equation}
with general solution
\begin{equation}\label{eq8}
\frac{1}{\alpha(\alpha-1)}\ln{\left[\frac{(H^2)^\alpha}{\Lambda_{UV}^2(\alpha-1)-\alpha H^2}\right]}=-\frac{2N}{\beta}+C,
\end{equation}
where $C$ is an integration constant and the function $H^2=H^2(N)$ is defined in an implicit way by this equation. It is noticeable that, for special values of $\alpha$ in Eq.~(\ref{eq7}), it is possible to obtain explicit analytical solutions for $H^2(N)$, for instance:
 \begin{equation}\label{eq9}
H^2(N)=\frac{1}{2}\left[e^{-\frac{N}{\beta}+C}-2\Lambda_{UV}^2\pm e^{-\frac{N}{\beta}}
\sqrt{e^{2C}-4e^{\frac{N}{\beta}+C}\Lambda_{UV}^2}\right]\ \ \ \ \left( \alpha=\frac{1}{2}\right),
\end{equation}
 \begin{equation}\label{eq10}
H^2(N)=-\frac{\Lambda_{UV}^2}{\text{ProductLog}\left[-e^{\frac{2N}{\beta}-C}\Lambda_{UV}^2\right]}\ \ \ \ \left(\alpha=1\right),
\end{equation}
and
\begin{equation}\label{eq11}
H^2(N)=\frac{1}{4}e^{-\frac{4N}{\beta}}\left[e^{2C}\pm\sqrt{e^{4C}-4e^{\frac{4N}{\beta}+2C}\Lambda_{UV}^2}\right]\ \ \ \ \left(\alpha=2\right),
\end{equation}
where ``ProductLog'' represents the Lambert W-Function.The general solution of Eq.~(\ref{eq6}) as function of $t$ is
\begin{equation}\label{equa}
\frac{1}{H(t)(1-\alpha)}+\frac{1}{\Lambda_{UV}\sqrt{(\alpha-1)^{3}\alpha}}\tanh^{-1}\left[\frac{H(t)\sqrt{\alpha}}{\Lambda_{UV}\sqrt{\alpha-1}}\right]
=-\frac{t}{\beta}+C.
\end{equation}
Again, as for $H(N)$, the function $H(t)$ is defined in an implicit way by Eq.~(\ref{equa}).\\

Usually, to analyze the dynamics of inflation in the slow-roll approximation,  it is a common practice to use the slow-roll parameters $\epsilon_n$ \cite{odintsov2}, which are given by
\begin{equation}\label{eq12}
\epsilon_{n+1}\equiv\frac{d\ln{\left|\epsilon_{n}\right|}}{dN},
\end{equation}
where $n$ is a positive integer and $\epsilon_0\equiv H_i/H$. Inflation terminates when the slow-roll parameter $\epsilon_1=1$ 
and the universe starts to reheating. Now, the values of the inflationary observables, i.e., 
the scalar spectral index of the curvature perturbations $n_s$, its running $\alpha_s \equiv dn_s/d\ln{k}$ (with $k$ the absolute
value of the wave number $\vec{k}$), the tensor spectral index $n_T$ and the tensor-to-scalar ratio $r$, may be characterized by the slow roll parameters $\epsilon_1$, $\epsilon_2$ and $\epsilon_3 $ \cite{martin}, as follows:
\begin{equation}\label{eq13}
n_s\approx 1-2\epsilon_1-2\epsilon_2,
\end{equation}
\begin{equation}\label{eq14}
\alpha_s\approx -2\epsilon_1\epsilon_2-2\epsilon_2\epsilon_3,
\end{equation}
\begin{equation}\label{eq15}
n_T\approx -2\epsilon_1\,,
\end{equation}
\begin{equation}\label{eq16}
r\approx 16\epsilon_1\,,
\end{equation}
and from Eq.~(\ref{eq12}) with $n=0,\ 1,\ 2$, it is straightforward to see that
\begin{equation}\label{eq17}
\epsilon_1=-\frac{H'}{H},
\end{equation}
\begin{equation}\label{eq18}
\epsilon_2=\frac{H''}{H'}-\frac{H'}{H},
\end{equation}
\begin{equation}\label{eq19}
\epsilon_3=-\frac{H'^4-HH'^2H''-H^2H''^2+H^2H'H'''}{HH'(H'^2-HH'')},
\end{equation}
where the prime denotes derivative with respect to e-folding variable $N$. Using Eq.~(\ref{eq7}) in the
last Eqs.~(\ref{eq17})-(\ref{eq19}), we get
\begin{equation}\label{eq20}
\epsilon_1=\frac{1}{\beta}\left(\alpha+\frac{1}{\tilde{H}^2-1}\right),
\end{equation}
\begin{equation}\label{eq21}
\epsilon_2=\frac{2\tilde{H}^2}{\beta(1-\tilde{H}^2)^2},
\end{equation}
\begin{equation}\label{eq22}
\epsilon_3=-\frac{2(1+\tilde{H}^2)\left[\alpha(1-\tilde{H}^2)-1\right]}{\beta(1-\tilde{H}^2)^2},
\end{equation}
where we have introduced the new function $\tilde{H}^2=H^2/\Lambda_{UV}^2$. Replacing Eqs.~(\ref{eq20})-(\ref{eq22}) in Eqs.~(\ref{eq13})-(\ref{eq16}) and eliminating
$\tilde{H}^2$, the inflationary observables are given by
\begin{equation}\label{eq23}
n_s=1+\frac{1}{8}r(4\alpha-3)-\frac{4\alpha(\alpha-1)}{\beta}-\frac{\beta r^2}{64},
\end{equation}
\begin{equation}\label{eq24}
\alpha_s=-\frac{r(\beta r-16\alpha)\left[16(1-\alpha)+\beta r\right]^2}{8192 \beta},
\end{equation}
and
\begin{equation}\label{eq25}
n_T=-\frac{r}{8}.
\end{equation}
We see that these inflationary observables are independent of  the ultraviolet cutoff $\Lambda_{UV}$. A similar
result was found in \cite{odintsov2}, but with a different choice for the infrared cutoff. We could also write the inflationary observables in terms of the e-folding variable $N$, using for example the explicit expressions for $H^2(N)$ given by Eqs. (\ref{eq9})-(\ref{eq11}).\\

\begin{figure}
\begin{center}
\includegraphics[scale=0.37]{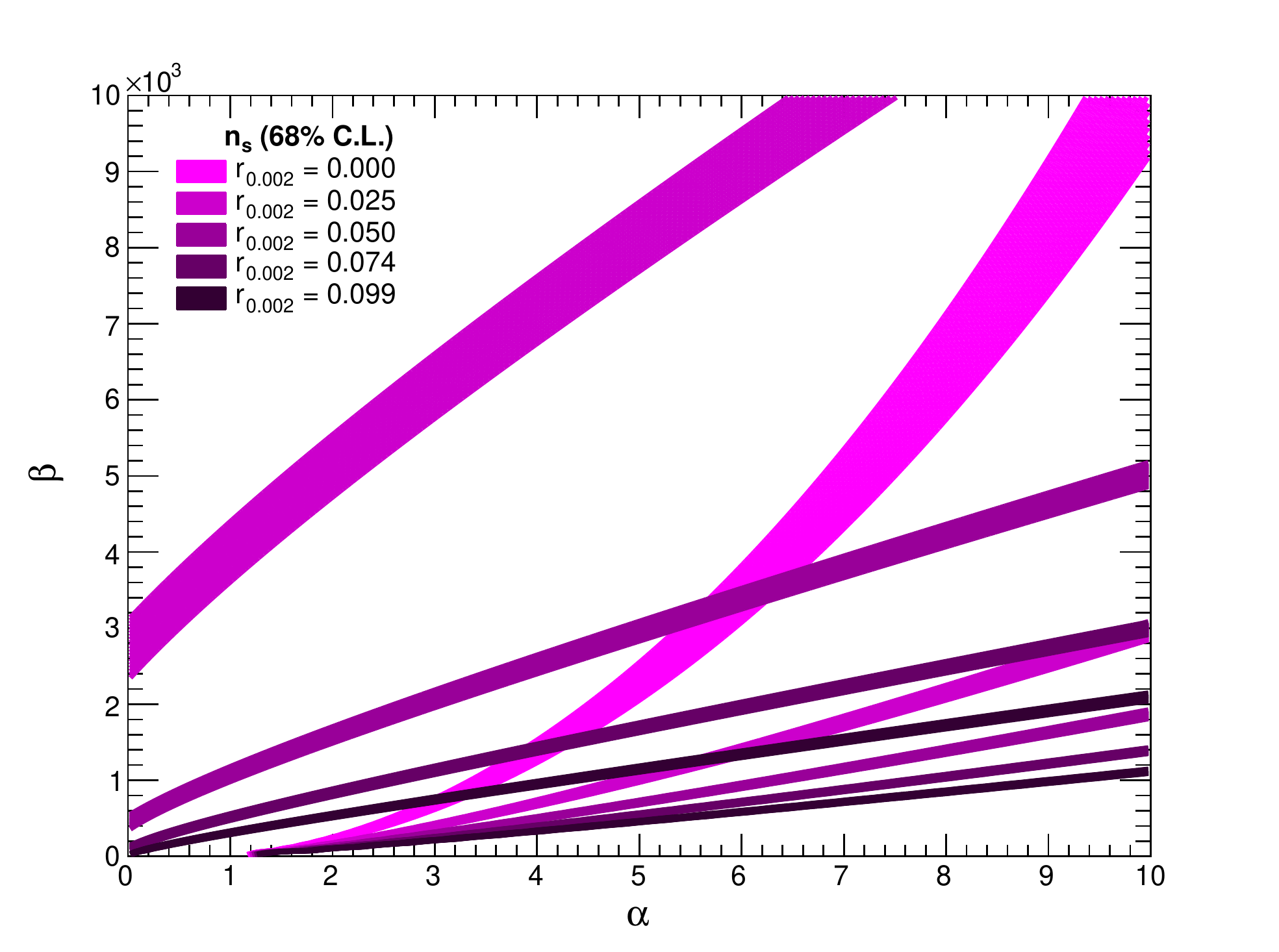} 
\includegraphics[scale=0.37]{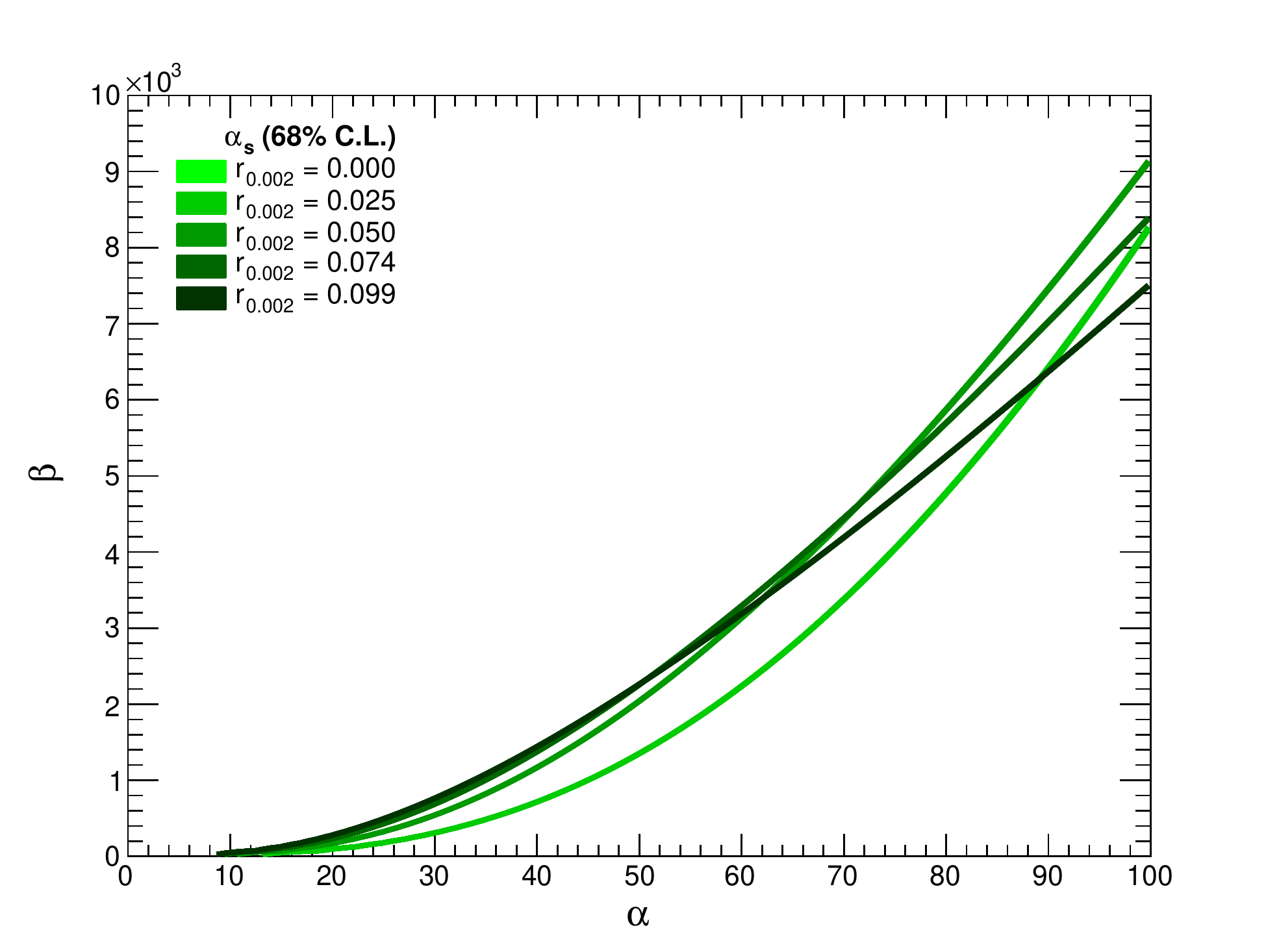}
\caption{Allowed values for $\alpha$ and $\beta$ parameters which define the model presented in this letter. Colored regions correspond to the allowed regions for $n_s$ (\emph{left}, purple) and $\alpha_s$ (\emph{right}, green) at the 68\% CL, for different values of $r$ inside the 95\% CL allowed region, according to Planck 2018 observations \cite{planck}. See text for details.}
\label{fig_contours}
\end{center}
\end{figure}

With Eq.~(\ref{eq23})-(\ref{eq25}) at hand,  we can investigate on the viability of our model, by looking for constraining the parameters $\alpha$ and $\beta$. We do this by analyzing the behavior of $n_s$ and $\alpha_s$ as functions of the parameters of the model, for different values of $r$ (within the allowed region found by the Plank Collaboration, at 95\% CL \cite{planck}). Figure \ref{fig_contours} shows the results of this analysis, where we plot the 68\% CL allowed values for $n_s$ (left) and $\alpha_s$ (right), for different values of $r$ in Eqs.~(\ref{eq24}) and (\ref{eq25}).\\

These plots should be read as follows: for any pair of the parameters ($\alpha$, $\beta$) inside the colored regions, one would reproduce allowed values for the scalar spectral index of the curvature perturbations $n_s$, its running $\alpha_s$ and the tensor-to-scalar ratio $r$. As one can notice, some of the colored regions in each figure overlap among each other, meaning that those values would give different valid combinations of $n_s$ and $r$ (purple regions), or  $\alpha_s$ and $r$ (green regions). To make this statement clear, for example, from the left panel of Figure \ref{fig_contours}, $(\alpha,\beta) = (6, 3400)$ are suitable values for the model parameters since with them one gets either $(n_s, r) = (0.9647, 0.0)$ or $(n_s, r) = (0.9631, 0.05)$, which are both well within the 68\% and 95\% CL allowed regions for $n_s$ and $r$, respectively \cite{planck}. Even more interesting is to notice that some of the green areas (right panel on Figure \ref{fig_contours}) also overlap with some of the purple ones. Such an overlap implies that the three observational parameters ($n_s$, $\alpha_s$, $r$) would be correctly reproduced by the model. This allows one to pose more stringent constraints on the values that $\alpha$ and $\beta$ should take to validate our model.\\

Overall, these plots are an understandable way to state the viability of our model by determining the posible values of the parameters which define it. The better the values of the inflationary observables are measured, the better the model parameter space is constrained, pointing to the values for $\alpha$ and $\beta$ which best fit to the observations. For instance, if it was proved that, say, $r < 0.8$, then $\alpha$ and $\beta$ figures inside the darkest regions on the plots of Figure \ref{fig_contours} would be excluded.\\

Finally, although for this model a covariant action which reproduces the Friedmann equations is still unknown, it may be equivalently described by some covariant theory (for example, modified gravity or fluid theory) \cite{odintsov1}. In this sense,
it is possible to obtain a correspondence between the G-O holographic energy density studied here and the effective density of the modified gravity
($F(R)$ and $F(G)$ gravity), which were performed  in \cite{granda} and \cite{oliveros3} in the dark energy context, but those could be used in an inflationary context too (for example, in \cite{paul} the author use the holographic correspondence with $F(R)$ gravity taking into account  extended Infrared cut-offs). Furthermore, in \cite{granda}, the Starobinsky $R^2$ term appears in some reconstructed solutions and it is well known that  the inflation based on Starobinsky proposal \cite{starobinsky} remains in good agreement with the current measurements of the cosmic microwave background \cite{planck}. Also, in \cite{oliveros2}, the authors have performed the correspondence between the G-O holographic energy density and some scalar field models of dark energy
(quintessence, tachyon, K-essence and dilaton). Additionally, this correspondence allows one to reconstruct the potentials and the dynamics for the scalar fields, which describe late accelerated expansion or, in our case, an inflationary  period. Likewise, again, it is well known that the scalar fields have played an important role in the explanation of the inflationary paradigm \cite{martin}. In view of such considerations, it is fair to say that the model presented here could lead to a covariant theory.

\section{Conclusions}
In this work we have investigated a model of inflation considering that the inflationary regimen has its origin in a 
type of holographic energy density. In particular, the Granda-Oliveros proposal for the holographic energy density
was considered (which has the so-called Granda-Oliveros infrared cutoff associated to it). Furthermore, since in the inflationary regimen the
energy scales are very high, it was necessary to modify the G-O cutoff by taking into account a correction due to the ultraviolet cutoff
$\Lambda_{UV}^2$. In this way, we found an algebraic equation which includes the Hubble parameter in an implicit form (see Eq.~(\ref{eq8})) and
from this, the Hubble slow-roll parameters (see Eqs. (\ref{eq20})-(\ref{eq22})) and the values of the inflationary observables $n_s$, $\alpha_s$, $n_T$ and $r$, were calculated (see Eqs. (\ref{eq23})-(\ref{eq25})). Additionally, using the allowed regions for these inflationary observables, we were able to pose constraints on model parameters, which are in agreement with Planck 2018 observations. Also, for this
holographic energy density used in this work, it's possible to obtain a correspondence between it and the effective density of the modified gravity ($F(R)$ and $F(G)$gravity) and also with some scalar fields models, which were shown in \cite{granda, oliveros2, oliveros3} in the dark energy context. Nevertheless, the formalism used in those references can be extended to an inflationary regime, too. Aside from this, since the parameters of the model studied here have a wide range of allowed values (see Fig.~\ref{fig_contours}), it is reasonable to think that, for some particular choice, the model would reproduce a holographic bounce in a similar fashion to the models studied in \cite{odintsov5, brevik}; however, that kind of analysis is beyond the scope of the present work and could be addressed later.

\end{document}